\documentclass[reprint,nofootinbib,superscriptaddress,amsmath,amssymb,aps]{revtex4-1}
\usepackage{graphicx}
\usepackage{rotating}
\usepackage{dcolumn}
\usepackage{bm}
\usepackage{color}
\usepackage{mathptmx, textcomp}
\usepackage[utf8]{inputenc}
\usepackage{braket}
\usepackage[colorlinks,linkcolor=magenta,anchorcolor=cyan,citecolor=blue]{hyperref}
\usepackage[multiple]{footmisc}

\def\be{\begin{equation}}
\def\ee{\end{equation}}
\def\ba{\begin{eqnarray}}
\def\ea{\end{eqnarray}}

      \def\p {\pi} \def\a {\alpha}  \def\d {\delta}            \def\m {\mu} \def\p {\pi}   
              \def\grad{\nabla}\def\.{\cdot}
\def\math {\mathcal}

\begin{document}

\title{Overcharging an accelerating Reissner-Nordstr\"{o}m-Anti-de Sitter black hole with test field and particle}
\author{Jie Jiang}
\email{jiejiang@mail.bnu.edu.cn}
\affiliation{College of Education for the Future, Beijing Normal University, Zhuhai 519087, China}
\author{Ming Zhang}
\email{mingzhang@jxnu.edu.cn (corresponding author)}
\affiliation{Department of Physics, Jiangxi Normal University, Nanchang 330022, China}

\date{\today}

\begin{abstract}
Accelerating black holes have been widely studied in the context of black hole thermodynamics, holographic gravity theories, and in the description of black holes at the center of galaxies. As a fundamental assumption to ensure spacetime causality, we investigated the weak cosmic censorship conjecture (WCCC) in the accelerating Reissner-Nordstr\"{o}m-Anti-de Sitter (RN-AdS) spacetime through the scattering of a charged field and the absorption of a charged particle. For the scattering of a charged scalar field, both near-extremal and extremal accelerating RN-AdS black holes cannot be overcharged, thereby upholding the validity of the WCCC. In the case of the absorption of a test charged particle, the results demonstrate that the event horizon of the extremal accelerating RN-AdS black hole cannot be destroyed, while the event horizon of the near-extremal black hole can be overcharged if the test particle satisfies certain conditions. The above results suggest that, in the case of test particles, second-order effects like self-force and self-energy should be further considered.

\end{abstract}
\maketitle

\section{Introduction}

The process of gravitational collapse, which inexorably leads to a spacetime singularity, is a consequence of the Hawking-Penrose singularity theorem \cite{Penrose:1964wq,Hawking:1970zqf}.  However, such singularities in gravitational theories may give rise to unpredictable outcomes, posing a significant challenge for scientists endeavoring to decipher universal laws. To address this issue, Penrose proposed the weak cosmic censorship conjecture (WCCC) \cite{Penrose:1969pc}, which aims to protect the predictability of gravitational theories. The conjecture states that naked singularities cannot be formed from regular initial data, and that singularities must be enveloped by an event horizon and remain hidden from distant observers. As such, the weak cosmic censorship conjecture is a fundamental open question in classical gravitational theory, and its resolution could provide important insights into the nature of the universe.

The weak cosmic censorship conjecture has been put through a myriad of tests including numerical simulations involving collapsing matter fields \cite{Christodoulou:1984mz,Ori:1987hg,Shapiro:1991zza,Lemos:1991uz,Choptuik:1992jv}, non-linear simulations of perturbed black holes or black rings \cite{Corelli:2021ikv,Eperon:2019viw,Crisford:2017zpi,Figueras:2017zwa,Figueras:2015hkb,Lehner:2010pn,Hertog:2003zs}, and simulations of merging black holes in higher dimensions \cite{Andrade:2020dgc,Andrade:2019edf,Andrade:2018yqu,Sperhake:2009jz}. Another methodology to probe the conjecture involves examining whether physical processes have the potential to destroy the event horizon. In 1974, Wald designed a gedanken experiment \cite{Wald:1974}, which demonstrated that an extremal Kerr-Newman black hole could withstand destruction from a test particle. However, the analysis was restricted to first-order perturbation and presupposed an extremal background black hole. Hubeny subsequently broadened the analysis to near-extremal Kerr-Newman black holes using second-order perturbation, showing that the near-extremal black hole could potentially be destroyed \cite{Hubeny:1998ga}. This assertion was supported by numerous subsequent studies \cite{deFelice:2001wj,Hod:2002pm,Jacobson:2010iu,Chirco:2010rq,Saa:2011wq,Gao:2012ca}. Yet, as Hubeny \cite{Hubeny:1998ga} discussed, to ascertain whether the black holes truly disintegrate, all second-order effects such as self-force and self-energy effects need to be taken into account. In 2017, Sorce and Wald \cite{Sorce:2017dst} proposed an updated version of the gedanken experiment, incorporating the full dynamical process of spacetime and colliding matter fields. After considering the null energy condition, their results suggested that a near-extremal Kerr-Newman black hole could not be destroyed under second-order perturbation. Following systematic investigations have reinforced the conclusion that the event horizon remains intact in these gedanken experiments across different gravitational systems \cite{An:2017phb,Ge:2017vun,Jiang:2019ige,Jiang:2019vww,Jiang:2019soz,Qu:2021hxh,Wang:2020vpn,Sang:2021xqj}.

In addition to particle injection, field scattering is another method of testing the WCCC. Superradiance, which refers to the phenomenon where a scalar field extracts energy from a charged or rotating black hole, results in interesting characteristics in the scattering of a scalar field. Semiz's work shows that an extremal dyonic Kerr-Newman black hole cannot be destroyed by a classical complex scalar field \cite{Semiz:2005gs}. Additionally, Gwak's research, which analyzes the scattering process in infinitesimal time intervals, demonstrated that extremal or near-extremal Kerr-(anti)de Sitter black holes cannot be overspun by test scalar fields \cite{Gwak:2018akg}. Studies of various other black holes also suggest that test scalar fields cannot destroy both extremal and near-extremal black holes \cite{Gwak:2021tcl,Liang:2020hjz,Natario:2016bay,Goncalves:2020ccm,Gwak:2019rcz,Yang:2020iat,Yang:2020czk,Feng:2020tyc,Yang:2022yvq}.

The traditional consensus among most studies has been that black holes are static or stationary entities, devoid of motion within space. Contrary to this assumption, numerous black holes have been discovered within binary systems. Gravitational waves emitted from these systems cause an increase in black hole velocity, making them seem as though they move and accelerate within our reference frame. Notably, accelerating black holes could also be generated by cosmic string fragmentation \cite{Hawking:1995zn,Eardley:1995au}, or within specific conditions such as a background magnetic field \cite{Garfinkle:1990eq,Hawking:1994ii,Hawking:1994ii}, or in de Sitter space \cite{Mellor:1989gi,Mann:1995vb,Dias:2003st,Ashoorioon:2014ipa,Ashoorioon:2020mos}. An alternate scenario involves a cosmic string network capturing primordial black holes during their formation \cite{Vilenkin:2018zol}, which would consequently lead to acceleration due to cosmic string tension. These accelerating black holes, if connected to cosmic strings, could evolve into supermassive black holes. However, if these are to participate in galaxy structure formation and take residence at galaxy centers, their acceleration needs to be small \cite{Vilenkin:2018zol,Morris:2017}. In light of this, Amjad Ashoorioon et al. have recently examined the properties of images in gravitational lensing effects within slowly accelerating black holes, proposing a novel method to observe black hole acceleration through gravitational lensing \cite{Ashoorioon:2022zgu,Ashoorioon:2021gjs}. The C-metric \cite{Weyl:1917} provides a key framework for describing accelerating black holes. This axisymmetric exact solution of field equations presents a boost symmetry \cite{Griffithsbook}. Its geometrical properties have been well established, yet the physical understanding of accelerating black holes remains incomplete, partly due to the previous lack of a suitable framework to study their thermodynamics. Recently, there have been further discussions on slowly accelerating black holes, particularly in the case of asymptotically Anti-de Sitter (AdS) spacetime \cite{Appels:2016uha,Gregory:2019dtq,Anabalon:2018qfv,Anabalon:2018ydc,Appels:2017xoe,Zhang:2019vpf}. Furthermore, the thermodynamic first law for accelerating black holes in asymptotically flat spacetime has also been addressed in Ref. \cite{Anabalon:2018qfv}. Within the context of general relativity, the C-metric has been employed to study radiation at infinity \cite{Ashtekar:1981ar,Podolsky:2003gm,Bicak:1988,Gomez:1994rg}, and the strong cosmic censorship conjecture \cite{Destounis:2020yav,Zhang:2023yco}. Moreover, its applications have found most resonance beyond the scope of classical general relativity, influencing studies on black hole pair creation \cite{Hawking:1995zn,Hawking:1997ia}, cosmic string fragmentation \cite{Eardley:1995au}, and notably, the construction of the black ring solution in five dimensions \cite{Emparan:2001wn}.

Given the fundamental necessity of the weak cosmic censorship conjecture in maintaining spacetime causality, it is of interest to ascertain if accelerating black holes align with it. Consequently, our paper aims to test the WCCC within the context of accelerating black holes, employing test fields and particles based on Semiz's method \cite{Gwak:2018akg,Gwak:2021tcl,Liang:2020hjz,Natario:2016bay,Goncalves:2020ccm,Gwak:2019rcz,Yang:2020iat,Yang:2020czk,Feng:2020tyc,Yang:2022yvq} in our analysis. From the previous literatures \cite{Gwak:2018akg,Gwak:2021tcl,Liang:2020hjz,Natario:2016bay,Goncalves:2020ccm,Gwak:2019rcz,Yang:2020iat,Yang:2020czk,Feng:2020tyc,Yang:2022yvq}, it is not hard to see that Semiz's approach is highly reliant on the thermodynamic first law of black holes. Given the challenges in thermodynamic studies of black holes for de Sitter accelerating black holes, and the absence of a fully established first law \cite{Appels:2016uha,Gregory:2019dtq,Anabalon:2018qfv,Anabalon:2018ydc,Appels:2017xoe,Zhang:2019vpf}, we only focus on the accelerating Reissner-Nordstr\"{o}m-Anti-de Sitter black hole in this paper. AdS spacetime is widely employed in research of the Anti-de Sitter/Conformal Field Theory (AdS/CFT) duality theory \cite{Maldacena:1997re,Gubser:1998bc,Witten:1998qj}. This theory establishes a profound connection between the gravitational theory in $D$-dimensional AdS spacetime and the CFT defined on its $(D - 1)$-dimensional boundary. Vacuum AdS spacetime corresponds to a zero-temperature state within the boundary CFT. On the other hand, an AdS black hole is associated with a thermal field double state on the boundary, with the temperature determined by the black hole's Hawking temperature \cite{Witten:1998zw}.  {If the black hole is overcharged, it results in the emergence of a naked singularity. Within the framework of AdS/CFT correspondence, naked singularities in the bulk gravitational system manifest as unconventional and unstable dual states in the boundary CFT. These dual states may exhibit a singular stress-energy tensor \cite{Chesler:2019ozd}. Additionally, these singularities may also lead to a phenomenon known as the "black tsunami'' in the boundary CFT \cite{Emparan:2021ewh}, characterized by non-thermal radiation bursts, indicating the high energy and dynamic instability during naked singularity formation.} Over the past two decades, AdS/CFT duality has seen widespread application in the fields of quantum chromodynamics and condensed matter theory.

The structure of this paper is organized as follows. In Sec. \ref{sec2}, we delve into an overview of the accelerating RN-AdS black hole and its associated thermodynamics. Sec. \ref{sec3} is dedicated to investigating the scattering of a charged scalar field within the context of an accelerating RN-AdS black hole background. We then turn our attention to the conserved charges for the charged scalar field during the scattering process in Sec. \ref{sec4}. In Secs. \ref{sec5} and \ref{sec6}, we make attempts to destroy the event horizon of the accelerating RN-AdS black hole utilizing a test scalar field and a test particle, respectively. Our concluding remarks and a summary of our findings are presented in the last section.

\section{Thermodynamics of slowly accelerating RN-AdS black hole\label{sec2}}

The slowly accelerating Reissner-Nordstrom-AdS black hole with a cosmological constant $\Lambda=-3/\ell^2$ is characterized by the metric and gauge potential, as detailed in \cite{Anabalon:2018qfv,Griffiths:2005qp}:
\begin{equation}\label{dsmet}\begin{aligned}
ds^2&=\frac{1}{\Omega^2}\left[-f(r)\frac{dt^2}{\alpha^2}+\frac{dr^2}{f(r)}+r^2\Big(\frac{d\theta^2}{h(\theta)}+h(\theta)\sin^2\theta\frac{d\phi^2}{K^2}\Big)\right]\\ \bm{F}&=d\bm B,\quad\quad \bm B=-\frac{e}{\alpha r}dt,
\end{aligned}\end{equation}
in which
\begin{equation}\begin{aligned}
f(r)&=(1-A^2r^2)\Big(1-\frac{2m}{r}+\frac{e^2}{r^2}\Big)+\frac{r^2}{\ell^2},\\
h(\theta)&=1+2mA\cos\theta+e^2A^2\cos^2\theta,\\
\alpha &= \sqrt{1+A^2 e^2-A^2\ell^2 (1+A^2 e^2)^2}.
\end{aligned}\end{equation}
The conformal factor is defined by
\begin{equation}\begin{aligned}
\Omega=1+A r\cos \theta.
\end{aligned}\end{equation}
This factor determines the conformal infinity, or the conformal boundary $r=r_b$ of the spacetime, where $r_b=-1/(A\cos\theta)$. $m$ and $e$ correspond to the black hole mass and electric charge parameters, respectively, while $A \geq 0$ is associated with the black hole's acceleration magnitude. The parameter $K$ induces the conical deficits at the poles. In the absence of $A$, the solution simplifies to the RN-AdS black hole.

The inner horizon $r=r_-$ and the event horizon $r = r_h$ of the accelerating RN black hole are ascertained from the equation $f(r)=0$. Beyond these two horizons, another horizon, called the accelerating horizon, is given by $r=r_A$, satisfying $r_A \geq r_h \geq r_-$. We also enforce a condition $1/A \geq r_A$ to ensure all horizons lie within the conformal boundary.

The black hole becomes extremal when $r_h = r_-$, i.e., both horizons coincide. Then, as the electric charge parameter $e$ increases, only one root $r=r_A$ of $f(r) = 0$ exists, causing the inner and event horizons to vanish, thereby depicting a naked singularity.

For the black holes, the mass $M$, electric charge $Q$, entropy $S$, thermodynamic pressure $P$, and tension $\mu_\pm$ at each pole $\theta\to \theta_\pm$
($\theta_+=0$ and $\theta_-=\pi$) are described as \cite{Appels:2017xoe,Anabalon:2018qfv}:
\ba\begin{aligned}\label{rl1}
M &= \frac{m \sqrt{1-A^2\ell^2 (1+A^2 e^2)}}{K\sqrt{1+A^2 e^2}},\quad Q=\frac{e}{K},\quad P=\frac{3}{8\pi \ell^2},\\
S&=\frac{\p r_h^2}{K(1-A^2 r_h^2)}, \quad \mu_\pm =\frac{1}{4}\left(1-\frac{1+A^2 e^2\pm 2m A}{K}\right)\,.
\end{aligned}\ea
The temperature $T$, electric potential $\Phi_H$, thermodynamic volume $V$, and thermodynamic potential $\lambda_\pm$ associated with the tension $\mu_\pm$ are given as \cite{Appels:2017xoe,Anabalon:2018qfv}:
\ba\begin{aligned}
 T &= \frac{f'(r_h)}{4\pi\alpha}\,,\quad \quad \quad\quad\Phi_H=\frac{e}{\a r_h},\\
 V&=\frac{4\p}{3K \alpha} \left[\frac{r_h^3}{(1-A^2 r_h^2)^2}+m A^2\ell^4(1+A^2 e^2)\right],\\
 \lambda_\pm &= \frac{Z_\pm}{2 \alpha  l^2 r_h \left(A^2 e^2+1\right) \left(A^2 r_h^2-1\right)}.
\end{aligned}\ea
Here, the additional parameters are:
\ba\begin{aligned}
Z_\pm=&\ell^2 r_h^2 \left(3 A^2 e^2+1\right)\mp2 A \ell^4 r_h \left(A^2 e^2+1\right)^2+r_h^4 \left(A^2 \ell^2-1\right)\\
&\pm 2 A \ell^2 r_h^3 \left(A^2 e^2+1\right) \left[A^2 \ell^2 \left(A^2 e^2+1\right)-1\right]-e^2 \ell^2.
\end{aligned}\ea
The first law of thermodynamics for this slowly accelerating RN-AdS black hole, which has been thoroughly studied, is expressed as \cite{Appels:2017xoe,Zhang:2019vpf}
\ba\begin{aligned}
\delta M = T \delta S+\Phi_H \delta Q+\lambda_+\delta\mu_++\lambda_-\delta\mu_-+V \delta P .
\end{aligned}\ea

\section{Massive complex scalar field in the accelerating RN-AdS spacetime}\label{sec3}

In this section, we turn our attention to the scattering of a massive complex scalar field $\varphi$, which is minimally coupled to gravity in the accelerating RN-AdS spacetime. The evolution of this complex scalar field is determined by the Klein-Gordon equation
\ba\begin{aligned}\label{fieldequation}
D_aD^a\phi -\mu_\text{s}^2\phi = 0.
\end{aligned}\ea
Here, the covariant derivative $D_a$ is defined as $D_a\equiv\nabla_a -i q B_a$, and $\mu_{\mathrm{s}}$ represents the mass of the complex scalar field $\varphi$. We can decompose the complex scalar field without loss of generality as
\ba\label{decom}
\varphi(t, r, \theta, \phi)=\mathrm{e}^{-\mathrm{i} \omega t} \mathrm{e}^{\mathrm{i} \bar m \phi} \psi(r, \theta).
\ea
By incorporating the metric (\ref{dsmet}) and the decomposition (\ref{decom}) into the motion equation (\ref{fieldequation}), we obtain:
\ba\begin{aligned}\label{eom2}
\frac{\partial^2\psi(r, \theta)}{\partial x^2}+\left(\alpha \omega -\frac{e q}{r}\right)^2\psi(r, \theta)+f(r)\hat{U} \psi(r, \theta)=0,
\end{aligned}\ea
in which $\hat{U}$ is an operator composed of the coordinates $r, \theta$, and $x$ is the tortoise coordinate defined by
\ba
\frac{\mathrm{d} r}{\mathrm{d} x}= f(r).
\ea
Near the event horizon, where $f(r) \rightarrow 0$, the equation (\ref{eom2}) simplifies to:
\ba\begin{aligned}
\frac{\partial^2\psi(r, \theta)}{\partial x^2}+\alpha^2\left( \omega -q \Phi_H\right)^2\psi(r, \theta)=0.
\end{aligned}\ea
Thus, the solution near the horizon can be written as
\ba\begin{aligned}
\psi(r, \theta)\propto e^{ \pm i \alpha (\omega-q \Phi_H) x}.
\end{aligned}\ea
The solution bifurcates into two branches: the first, characterized by a positive sign, represents an outgoing wave, while the second, denoted by a negative sign, signifies an ingoing wave. For the purposes of our study, we select the latter as it aligns with physically acceptable solutions. Consequently, the wave function near the horizon becomes
\ba\begin{aligned}\label{ingoingpsi}
\varphi(t, r, \theta, \phi)=\mathrm{e}^{-\mathrm{i} \omega t+i \bar m \phi} \mathrm{e}^{-i\alpha\left(\omega-q \Phi_H\right) x} \Theta(\theta).
\end{aligned}\ea
In this equation, $\Theta$ is a function of $\theta$, derived from the field equation (\ref{eom2}) near the horizon, and we assume it to be compact on the horizon's cross section. Following this, our aim lies in investigating the changes to the black hole parameters after the scattering of the ingoing wave, using the above wave function as a reference.

\section{Conserved charges under scattering of the scalar field}\label{sec4}

In the context of our study, we disregard the influence of self-force and other interactions, implying that the energy and electric charge conveyed by the wave are sufficiently small. The variation in black hole energy aligns with the energy flux, which is established by the energy-momentum tensor of the massive scalar field
\ba\begin{aligned}\label{Tab}
T_{b}^{a}=&\frac{1}{2} D^{a} \varphi \partial_{b} \varphi^{*}+\frac{1}{2} D^{* a} \varphi^{*} \partial_{b} \varphi\\
&-\delta_{b}^{a}\left(\frac{1}{2} D^{c} \varphi D_{c}^{*} \varphi^{*}+\mu_{s}^{2} \varphi^{*} \varphi\right).
\end{aligned}\ea
Employing the energy-momentum tensor (\ref{Tab}) along with the ingoing wave function (\ref{ingoingpsi}), the energy flux traversing the event horizon is computed by
\ba\begin{aligned}\label{dEdt}
\frac{\mathrm{d} E}{\mathrm{~d} t}=\int_{H} T_{t}^{r} \sqrt{-g} \mathrm{~d} \theta \mathrm{d} \phi=\omega\left(\omega-q \Phi_H\right)\math{B},
\end{aligned}\ea
and the charge flux is
\ba\begin{aligned}\label{dQdt}
\frac{\mathrm{d} Q}{\mathrm{~d} t}=\int_{H} j^{r} \sqrt{-g} \mathrm{~d} \theta \mathrm{d} \phi=q\left(\omega-q \Phi_H\right)\math{B}.
\end{aligned}\ea
Here, the electric current $j^{\mu}$ is given by
\ba\label{ja}
j^{a}=-\frac{1}{2} i q\left[\varphi^{*}D^a \varphi-\varphi D^{*a} \varphi^{*}\right],
\ea
and we have defined
\ba\begin{aligned}
\math{B}=2\pi \int_{0}^{\pi}\frac{r_h^2\sin \theta |\Theta(\theta)|^2}{K \Omega^2}.
\end{aligned}\ea
With the energy flux (\ref{dEdt}) and the charge flux (\ref{dQdt}), we can get the changed energy and charge within a given infinitesimal time interval $\delta t$ as
\ba\begin{aligned}\label{dEdQ}
\delta M&= d E=\omega\left(\omega-q \Phi_H\right) \math{B}\delta t,\\
\delta Q&= q\left(\omega-q \Phi_H\right) \math{B} \delta t.
\end{aligned}\ea

{  In the above calculation, we adopted a specific gauge choice where $A_a$ tends to zero as $r$ approaches infinity. This choice differs from the one utilized in the solution presented in Ref. \cite{Anabalon:2018qfv}. The rationale behind this particular gauge choice is our requirement that the energy-momentum tensor of the scalar field remains independent of the electromagnetic field at infinity. This condition is motivated by the fact that the electromagnetic strength tensor $F_{ab}$ vanishes as we approach infinity.}

From the above derivations, it's clear that the relationship between $\omega$ and $q \Phi_H$ governs the signs of the energy flux and the charge flux. The energy and charge of the black hole increase when $\omega>q \Phi_H$, remain constant when $\omega=q \Phi_H$ and decrease when $\omega<q \Phi_H$. This decreasing state implies that the energy and charge are siphoned off by the scattering field, a phenomenon referred to as superradiance \cite{R. Brito:2015}. { Semiz and Duztas thoroughly explored the relationship between superradiance and quantum particle creation in the context of the WCCC in their research \cite{Semiz:2015pna}. Their study revealed that superradiance may play a significant role within the framework of the WCCC and could potentially prevent black holes from becoming overcharged.}

The accelerating RN-AdS black hole is described by five parameters: the mass parameter $m$, the electric parameter $e$, the acceleration $A$, the deficit parameter $K$, and the cosmological constant $\Lambda$. A natural question arises as to which of these parameters will change during the scattering process. Next, we will follow the approach outlined in Refs. \cite{Feng:2020tyc,Yang:2020iat}, and apply the laws of black hole thermodynamics to investigate the variations of the black hole parameters during the scattering process.

It's important to note that the cosmological constant stays fixed during scattering as it is a theoretical parameter. Assuming that the test field only influences the energy and electric charge of the black hole, and that the acceleration $A$ and deficit parameter $K$ are fixed during this process, we can apply the first law of black hole thermodynamics to obtain
\ba
\begin{aligned}
\delta S & =\frac{1}{T}\left(\delta M-\Phi_H \delta Q-\lambda_+\delta\mu_+-\lambda_-\delta\mu_-+V \delta P\right) \\
& =\frac{q^{2} A(1+A^2e^2)}{\alpha^2(1-A^2r_h^2)} \left(\frac{\omega}{q}-\Phi_H\right)\left(\frac{\omega}{q}-\tilde{\Phi}\right)\delta t,
\end{aligned}
\ea
where
\ba\begin{aligned}
\tilde{\Phi}=\frac{e Y}{2 \alpha \ell^2 r_h \left(A^2 e^2+1\right) \left(A^2 r_h^2-1\right)}
\end{aligned}\ea
with
\ba\begin{aligned}
Y=&\ell^2 \left(A^2 r_h^2-1\right) \left[A^2 \left(e^2-r_h^2\right)+2\right]\\
&+A^2 r_h^4-2 \ell^4 \left(A^3 e^2+A\right)^2 \left(A^2 r_h^2-1\right).
\end{aligned}\ea
For a slowly accelerating black hole with a small value of $A$, we derive
\ba\begin{aligned}
\tilde{\Phi} = \Phi_H-\frac{A^2 e \left(e^2 \ell^2+2 \ell^4+\ell^2 r_h^2+r_h^4\right)}{2 \ell^2 r_h}+\mathcal{O}(A^3),
\end{aligned}\ea
which infers that
\ba
\Phi_H >\tilde{\Phi}
\ea
for slowly accelerating scenarios. For wave modes with $\omega / q$ satisfying
\ba
\Phi_H>\frac{\omega}{q}>\tilde{\Phi},
\ea
the entropy $S$ of the black hole decreases during the scattering process, contradicting the second law of black hole thermodynamics. Consequently, the parameters $A$ and $K$ must undergo changes during the scattering process.

Let's now consider a scenario where the tensions $\mu_\pm$ remain constant during the scattering process. The entropy change is then given by
\ba\begin{aligned}
\delta S & =\frac{1}{T}\left(\delta M-\Phi_H \delta Q-\lambda_+\delta\mu_+-\lambda_-\delta\mu_-+V \delta P\right) \\
& =\frac{q^{2} A(1+A^2e^2)}{\alpha^2(1-A^2r_h^2)} \left(\frac{\omega}{q}-\Phi_H\right)^2\delta t .
\end{aligned}\ea
This result indicates that the entropy never decreases, thus adhering to the second law of black hole thermodynamics. Therefore, we will adopt the assumption that the tensions $\mu_\pm$ remain fixed under the scattering process in the subsequent discussions.

\section{Destroying the event horizon with test scalar field}\label{sec5}

In this section, we aim to destroy the event horizon of an extremal and a near-extremal accelerating RN-AdS black hole by scattering a classical complex scalar field into them. As a result of this scattering process, the energy and charge of the black hole undergo changes. To ensure the existence of the event horizon, we need the minimum value of the metric function $\Delta(r) = r^2 f(r)$ to be non-positive. By examining this condition, we can determine if the black hole is destroyed. For an accelerating RN-AdS black hole, the minimum value $\Delta_\text{min}$ of the metric function $\Delta(r)$ is given by:
\ba\begin{aligned}
\Delta_\text {min }&\equiv\Delta\left(r_{\min }\right)\\
& =(1-A^2r^2_\text{min})\Big(r_\text{min}^2-2m r_\text{min}+e^2\Big)+\frac{r^4_\text{min}}{\ell^2},
\end{aligned}\ea
where $r_\text{min}$ is determined by
\ba\begin{aligned}
\Delta'\left(r_\text{min}\right)=&m \left(6 A^2 r_\text{min}^2-2\right)\\
&+r_\text{min} \left[\frac{4 r_\text{min}^2}{\ell^2}+2-2 A^2 \left(e^2+2 r_\text{min}^2\right)\right]=0.
\end{aligned}\ea
After substituting $m, e, K, A, \ell$ with $M, Q, \mu_\pm, P$ using the relationships \eqref{rl1}, the metric function $\Delta(r)$ can be expressed as a function of $r, M, Q, \mu_\pm$, and $P$, denoted as $\Delta=\Delta(r, M, Q, \mu_\pm, P)$.

After the scattering of the scalar field, the parameters of the final state are changed as
\ba\begin{aligned}
M &\rightarrow M^{\prime}  =M+\mathrm{d} M, \quad Q  \rightarrow Q^{\prime}=Q+\mathrm{d} Q, \\
\mu_\pm &\rightarrow \mu_\pm^{\prime}  = \mu_\pm,\quad\quad\quad P \rightarrow P^{\prime}  = P.
\end{aligned}\ea
In the previous section, we discussed that the tensions $\mu_\pm$ and the pressure $P$ remain unchanged during the scattering process. Now, let's consider the minimum value of the metric function $\Delta(r, M+\delta M, Q+\delta Q, \mu_\pm, P)$ in terms of the initial state function $\Delta(r_\text{min}, M, Q, \mu_\pm, P)$. In the case of a naked singularity, the function $\Delta(r, M+\delta M, Q+\delta Q, \mu_\pm, P)$ has a positive minimum value due to overcharging. As a result of the scattering of the scalar field, the minimum point infinitesimally shifts to $r_{\text{min}}+\delta r_{\text{min}}$. For these infinitesimal changes, the minimum value of the function $\Delta$ can be expressed as:
\ba\begin{aligned}\label{overchargecd1}
&\Delta\left(r_\text{min}+\delta r_\text{min}, M+\delta M, Q+\delta Q, \mu_\pm, P\right)\\
&=\Delta_\text{min}+\frac{\partial \Delta}{\partial M}\delta M+\frac{\partial \Delta}{\partial Q}\delta Q,
\end{aligned}\ea
where $\Delta_\text{min}$ specifically denotes the minimum value of the initial state, and
\ba
\begin{aligned}
& \frac{\partial \Delta}{\partial M} \equiv \left(\frac{\partial \Delta}{\partial M}\right)_{r=r_\text{min}, Q, \mu_\pm, P},\\
& \frac{\partial \Delta}{\partial Q} \equiv \left(\frac{\partial \Delta}{\partial Q}\right)_{r=r_\text{min}, M, \mu_\pm, P}.
\end{aligned}
\ea

By considering the energy and electric charge fluxes, as well as the initial conditions in Eqs. (\ref{dEdQ}) and (\ref{overchargecd1}), we can determine the minimum value after scattering as
\ba
\begin{aligned}
&\Delta\left(r_\text{min}+\delta r_\text{min}, M+\delta M, Q+\delta Q, \mu_\pm, P\right)\\
& =\Delta_\text{min}+q^{2} \math{A} \frac{\partial \Delta_r}{\partial M}\left(\frac{\omega}{q}-\Phi_H\right)\left(\frac{\omega}{q}-\Phi_{\mathrm{eff}}\right) \delta t,
\end{aligned}
\ea
where the effective electric potential $\Phi_{\text{eff}}$ plays a crucial role in determining the sign of the minimum value in the final state. It is defined as
\ba
\begin{aligned}
\Phi_{\mathrm{eff}} \equiv-\left(\frac{\partial_Q \Delta}{\partial_M \Delta}\right).
\end{aligned}\ea
Since $\Phi_{\text{eff}}$ is expressed in terms of $r_{\text{min}}$, we need to rewrite $\Phi_{\text{eff}}$ in terms of $r_h$ for comparison with $\Phi_H$. For near-extremal black hole solutions, the outer horizon is extremely close to the minimum point. Assuming an infinitesimal distance $\epsilon$ between the minimum point and the outer horizon, i.e.,
\ba\begin{aligned}
r_\text{min}=r_h(1+\epsilon),
\end{aligned}\ea
we can express the initial minimum value $\Delta_{\text{min}}$ in terms of $\epsilon$. Using the conditions $\Delta(r_h, M,\cdots)=0$ and $\partial_r \Delta(r_{\text{min}}, M,\cdots)=0$, we find
\ba\begin{aligned}
\Delta_\text{min}=\epsilon^2 X+\mathcal{O}(\epsilon^3),
\end{aligned}\ea
where
\ba\begin{aligned}
X = r_h^2\left(\frac{2 \left(A^2 e^2-3\right)}{A^2 r_h^2-3}+ A^2 e^2-\frac{2 e^2}{r_h^2}+1\right) <0,
\end{aligned}\ea
and $\epsilon \geq 0$. Furthermore, we can also obtain
\ba\begin{aligned}
\frac{\partial \Delta}{\partial M}=&-2K \alpha r_h (1-A^2r_h^2)^2\\
&+2\epsilon K\alpha r_h (1-A^2r_h^2)(1-5 A^2 r_h^2)+\mathcal{O}(\epsilon^2),\\
\frac{\partial \Delta}{\partial Q}=&2 e K (1-A^2r_h^2)^2+8\epsilon K A^2 e r_h (1-A^2r_h^2)+\mathcal{O}(\epsilon^2).
\end{aligned}\ea
The above result indicates that $(\partial_M \Delta) <0$ for slowly accelerating RN-AdS black holes. Consequently, the effective electric potential $\Phi_{\mathrm{eff}}$ can be expressed in terms of $r_{\mathrm{h}}$ and $\epsilon$ as follows:
\ba\begin{aligned}\label{Phieff}
\Phi_{\mathrm{eff}}=\Phi_H(1+\epsilon)+\mathcal{O}\left(\epsilon^{2}\right),
\end{aligned}\ea
where the signs of $\Phi_{\mathrm{eff}}$ and $\Phi_H$ coincide. In the case of a general cosmological constant $\Lambda$ and acceleration $A$, the effective electric potential is slightly greater than $\Phi_H$ for $e>0$. The overcharging process occurs exclusively when $e>0$ and $q>0$. Hence, without loss of generality, we assume $e>0$ and $q>0$. Then, the destruction condition $\Delta\left(r_\text{min}+\delta r_\text{min}, M+\delta M, Q+\delta Q, \mu_\pm, P\right)>0$ reduces to
\ba\begin{aligned}\label{judgeeq1}
\left(\frac{\omega}{q}\right)^{2}&-\left(\Phi_{\mathrm{eff}}+\Phi_H\right)\left(\frac{\omega}{q}\right)+\Phi_{\mathrm{eff}}\Phi_H\\
&+\frac{\Delta_\text{min}}{q^{2} \math{A} (\partial_M \Delta)\delta t}<0,
\end{aligned}\ea
where we have used $(\partial_M \Delta) <0$. Here, we assume that two parameters, $\epsilon$ and $\delta t$, are infinitesimal in scale. In practice, we can freely define $\epsilon$ such that $\epsilon \sim \delta t$. The inequality (\ref{judgeeq1}) has no solution because the discriminant is given by:
\ba
-\frac{4X \epsilon}{q^{2} \math{A} (\partial_M \Delta)}+\Phi_H^2 \epsilon^2+\mathcal{O}(\epsilon^3),
\ea
which is negative to first order in $\epsilon$. Therefore, during the scattering process of the scalar field, the near-extremal accelerating RN-AdS black hole cannot be overcharged. Furthermore, the final black hole becomes more non-extremal than the initial one for any combination of $\omega$ and $q$ of the scalar field. In other words, the energy transferred from the scalar field is greater than the electric charge transferred from it. For the case of equality $(\omega / q)=\Phi_H$ or $(\omega / q)=\Phi_{\mathrm{eff}}$, the change in the minimum value $\Delta_\text{min}$ also becomes zero, resulting in the initial and final states being identical. {Furthermore, it's worth noting that throughout these calculations, we have not imposed any constraints on the $\omega/q$ value. Hence, scalar fields of any $\omega/q$ do not lead to black hole overcharging, even in cases exhibiting superradiance (where $\omega/q < \Phi_H$).}

\section{Destroy the event horizon with test particle}\label{sec6}

One other approach to potentially destroy the event horizon of an accelerating RN-AdS black hole is by dropping a test charged particle into it. This gedanken experiment, first proposed by Wald, explores the possibility of destroying the black hole's event horizon. By neglecting the back-reaction, Hubeny \cite{Hubeny:1998ga} demonstrated that a near-extremal Reissner-Nordstrom black hole can become overcharged after capturing a test particle. In this section, we aim to investigate whether the event horizon of an accelerating RN-AdS black hole can still persist after absorbing a test charged particle.

The Lagrangian of the test particle, characterized by its rest mass $\mu_m$ and charge $\delta Q$, is given by
\ba\label{latestpartical}
L=\frac{1}{2} \mu_m U^a U_a+\delta Q U^a B_a,
\ea
where
\ba\begin{aligned}
U^a=\left(\frac{\partial}{\partial \tau}\right)^a
\end{aligned}\ea
represents the four-velocity of the particle and $\tau$ denotes the particle's proper time, ensuring $U^a U_a = -1$. By deriving the equation of motion from the Lagrangian (\ref{latestpartical}), we obtain
\ba\begin{aligned}
U^b\grad_b U^a=\frac{\delta Q}{\mu_{m}} F^{a}{ }_{b} U^b.
\end{aligned}\ea

In this study, we focus on the scenario where the particle is dropped onto the equatorial plane $(\theta=\pi / 2)$ without any angular momentum, resulting in vanishing components of angular momentum $P_{\phi}$ and $P_{\theta}$. The energy and angular momentum are determined as
\ba\begin{aligned}
\delta E &=-\frac{\partial L}{\partial \dot{t}}=-\mu_m U_t-\delta Q B_{t},\\
P_{\phi} &=\frac{\partial L}{\partial \dot{\phi}}=\mu_m U_\phi=0,\\
P_{\theta} &=\frac{\partial L}{\partial \dot{\theta}}=\mu_m U_\theta=0.
\end{aligned}\ea

Next, we explore the conditions required to destroy the black hole. Specifically, the test particle must be capable of entering the event horizon, and the black hole should become overcharged after absorbing the test particle. These conditions establish a relationship between the energy $\delta E$ and charge $\delta Q$ of the test particle.

The condition for the test particle to enter the event horizon requires that its motion outside the event horizon is timelike and future-directed, given by
\ba\begin{aligned}
U^aU_a = g_{\mu \nu} \frac{\mathrm{d} x^{\mu}}{\mathrm{d} \tau} \frac{\mathrm{d} x^{\nu}}{\mathrm{d} \tau}&=-1,\quad\text{and}\quad \frac{d t}{d \tau} > 0 .
\end{aligned}\ea
Using the above results, we obtain
\ba\begin{aligned}
\frac{d t}{d\tau}&=\frac{\alpha \Omega^2 (\alpha r \d E-\d Q e)}{\m_m r f(r)},\\
\left(\frac{d r}{d\tau}\right)^2&=\frac{\Omega^2[\Omega^2(\alpha r \d E-\d Q e)^2-\mu_m^2 f(r)r^2]}{\mu_m^2 r^2}.
\end{aligned}\ea

The condition for the test particle to enter the event horizon can be obtained by requiring the trajectory of the charged particle outside the event horizon to be future-directed:
\ba
\delta E > \Phi_H \delta Q.
\ea

After the test particle is dropped into the black hole, the parameters of the final state become
\ba\begin{aligned}
M &\rightarrow M^{\prime}  =M+\mathrm{d} M = M+\delta E, \\
Q  \rightarrow Q^{\prime}=Q+&\delta Q, \quad \mu_\pm \rightarrow \mu_\pm^{\prime}  = \mu_\pm,\quad P \rightarrow P^{\prime}  = P.
\end{aligned}\ea
Furthermore, the condition for the black hole to be overcharged requires that the minimum value of the metric function $\Delta\left(r_\text{min}+\delta r_\text{min}, M+\delta M, Q+\delta Q, \mu_\pm, P\right)$ is positive. From Eq. (\ref{overchargecd1}), the minimal value of $\Delta$ after dropping the particle to first order is
\ba
\begin{aligned}
&\Delta\left(r_\text{min}+\delta r_\text{min}, M+\delta M, Q+\delta Q, \mu_\pm, P\right)\\
& =\Delta_\text{min}+\frac{\partial \Delta_r}{\partial M}\left(\delta M-\Phi_{\mathrm{eff}}\delta Q\right).
\end{aligned}
\ea

For an extremal accelerating RN-AdS black hole, we have $\Delta_{\min }=0$ and $r_m=r_h$. Then, we have $\Phi_\text{eff}=\Phi_H$, which implies
\ba
\begin{aligned}
&\Delta\left(r_\text{min}+\delta r_\text{min}, M+\delta M, Q+\delta Q, \mu_\pm, P\right)\\
& =\frac{\partial \Delta_r}{\partial M}\left(\delta M-\Phi_H\delta Q\right)\geq 0.
\end{aligned}
\ea
This means that the test charged particle, which can potentially destroy the event horizon, experiences a repulsive force from the black hole and cannot enter the event horizon. Thus, the event horizon of the extremal accelerating RN-AdS black holes cannot be overcharged.

For a near-extremal accelerating RN-AdS black hole with $r_\mathrm{min}=r_h(1-\epsilon)$, utilizing Eq. (\ref{Phieff}), we find that
\ba
\begin{aligned}
&\Delta\left(r_\text{min}+\delta r_\text{min}, M+\delta M, Q+\delta Q, \mu_\pm, P\right)\\
& =\frac{\partial \Delta}{\partial M}\left(\delta M-\Phi_H\delta Q\right)+\epsilon^2 X-\epsilon\frac{\partial \Delta}{\partial M}\Phi_H\delta Q+\mathcal{O}(\epsilon^3).
\end{aligned}
\ea
This outcome suggests that the event horizon of a near-extremal accelerating black hole can be disrupted by a charged test particle. To illustrate examples of the black hole destruction, we can choose test particle parameters that satisfy
\ba\begin{aligned}
\delta E = \Phi_H \delta Q,\quad\quad \delta Q >  \frac{\epsilon X}{\Phi_H\partial_M\Delta} >0,
\end{aligned}\ea
implying that
\ba
\begin{aligned}
&\Delta\left(r_\text{min}+\delta r_\text{min}, M+\delta M, Q+\delta Q, \mu_\pm, P\right)\\
& =\epsilon\Phi_H(\partial_M\Delta)\left(\frac{\epsilon X}{\Phi_H\partial_M\Delta}-\delta Q\right) > 0
\end{aligned}
\ea
under the second-order approximation of $\epsilon$. Here, we have employed the condition $(\partial_M \Delta) < 0$. The chosen values for the energy and electric charge of the test particle are reasonable since they are both small quantities of first-order magnitude in $\epsilon$, allowing the particle to be treated as a test particle. Consequently, the event horizon of a near-extremal accelerating RN-AdS black hole can be destroyed by certain charged particles.

\section{Conclusion and discussion}\label{sec7}

In this paper, we investigated the possibility of destroying the event horizon of the accelerating RN-AdS black hole using a test charged scalar field and a test charged particle, separately. By applying the thermodynamic laws of the accelerating RN-AdS black hole, we showed that the tensions $\mu_\pm$ remain unchanged during the scattering process, while the acceleration $A$ and deficit parameter $K$ need to change. In the case of the test charged scalar field scattering, we found that both extremal and near-extremal accelerating RN-AdS black holes cannot be overcharged. The energy and charge fluxes from the scalar field do not lead to the destruction of the event horizon, thus supporting the WCCC. On the other hand, when considering the absorption of a test charged particle, we discovered that the event horizon of an extremal accelerating RN-AdS black hole remains intact. However, for a near-extremal black hole, the event horizon can be destroyed if the test particle satisfies certain conditions. In such a case, the black hole becomes overcharged and loses its event horizon, violating the WCCC. { Such occurrences are not surprising, primarily due to the omission of self-force and self-energy effects, as similar situations have been observed in the context of Kerr-Newman black holes \cite{Hubeny:1998ga}. Consequently, these results suggest the need for further consideration of the impact of self-force and self-energy on the validity of the WCCC.

However, as noted by Sorce and Wald, the analytical computation of electromagnetic and gravitational self-force effects of the test body is currently beyond our capabilities even in the Kerr-Newman black holes. Fortunately, they have introduced a new gedanken experiment using the Noether charge method to examine the WCCC, which automatically incorporates second-order effects, including self-force and self-energy. Nevertheless, it is essential to recognize that due to the presence of conical defects, and the departure from asymptotic flatness, certain aspects of their computations may not straightforwardly be applicable to accelerating black holes. Therefore, we plan to leave the extension of their computational techniques to future research.}

{ Due to the strong reliance of the Semiz's method on the first law of black hole thermodynamics and the absence of a fully established first law for de Sitter accelerating black holes, we only focus our study exclusively on the asymptotically Anti-de Sitter case. In future research, after conducting a more comprehensive exploration of black hole thermodynamics in de Sitter spacetimes, it will be essential to further investigate the WCCC.} Furthermore, it is also meaningful to investigate the case of accelerating black holes in the presence of rotation. Additionally, studying other types of test fields and particles, such as fermionic fields, is also crucial for testing the WCCC.

\section*{Acknowledgements}
J. J. is supported by the National Natural Science Foundation of China with Grant No. 12205014, the
Guangdong Basic and Applied Research Foundation with Grant No. 2021A1515110913, and the Talents Introduction
Foundation of Beijing Normal University with Grant No. 310432102. M. Z. is supported by the National Natural Science Foundation of China with Grant No. 12005080.

\end{document}